\documentclass[conference]{IEEEtran}
\IEEEoverridecommandlockouts

\newcommand{\conftitle}{17th International Workshop on Science Gateways (IWSG2025), 17-19 June 2025}

\usepackage{fancyhdr}
\usepackage{url}
\usepackage{float}
\usepackage{xcolor}
\usepackage{subcaption}
\fancypagestyle{pageStyle}{%
    \fancyhf{}
    \fancyhead[C]{\conftitle}

}


\ifCLASSINFOpdf
  \usepackage[pdftex]{graphicx}
\else
\fi

\usepackage{listings}%
\usepackage[most]{tcolorbox}%

\newif\ifproofread

\newcommand{\changemarker}[1]{%
\ifproofread
\textcolor{blue}{#1}%
\else
#1%
\fi
}

\proofreadfalse

\begin{document}

%
\title{Developing a Portable Solution for Post-Event Analysis Pipelines}
\renewcommand\IEEEkeywordsname{Keywords}



%
%

%

\author{\IEEEauthorblockN{Leonardo Pelonero, Fabio Vitello, Eva Sciacca, Mauro Imbrosciano, Salvatore Scavo, Ugo Becciani
}
\IEEEauthorblockA{INAF Astrophysical Observatory of Catania, Via Santa Sofia 78, Catania, Italy}
Email: leonardo.pelonero@inaf.it}



\maketitle
\thispagestyle{pageStyle}
\pagestyle{fancy}
\renewcommand{\headrulewidth}{0pt} 

\begin{abstract}


In recent years, the monitoring and study of natural hazards have gained significant attention, particularly due to climate change, which exacerbates incidents like floods, droughts, storm surges, and landslides. Together with the constant risk of earthquakes, these climate-induced events highlight the critical necessity for enhanced risk assessment and mitigation strategies in susceptible areas such as Italy.

In this work, we present a Science Gateway framework for the development of portable and fully automated 
post-event analysis pipelines integrating
Photogrammetry techniques, Data Visualization and Artificial Intelligence technologies, applied on aerial images, to assess extreme natural events and evaluate their impact on risk-exposed assets.
\end{abstract}


%
\IEEEpeerreviewmaketitle

\section{Introduction}


The monitoring and analysis of natural hazards have gained increasing relevance in recent years, as their frequency and intensity continue to rise, particularly under the influence of climate change, which intensifies the frequency and severity of events such as floods, droughts, storm surges, and landslides. These climate-driven phenomena highlight the urgent need for more effective risk assessment and mitigation strategies, especially in highly vulnerable areas like Italy.

Hazard mapping plays a central role in this context, as it enables the identification and spatial representation of risk-prone areas. Having a ready-to-use and easily deployable analysis pipeline is of paramount importance when responding to disastrous events. To address this need, automated pipelines that orchestrate photogrammetric workflows and AI-driven image preprocessing are essential to support timely decision-making.


The National Institute for Astrophysics (INAF), together with other national institutions, is partner in the HaMMon (Hazard Mapping and vulnerability Monitoring) project — an initiative launched within the Italian National Research Centre for High Performance Computing, Big Data and Quantum Computing (ICSC)\footnote{ICSC: \url{https://www.supercomputing-icsc.it/}} and coordinated by UnipolSai\footnote{UnipolSai: \url{https://www.unipolsai.it/}}. The project aims to develop advanced tools and methodologies for the management of natural hazards, addressing all aspects from risk assessment to post-event analysis, including intervention planning and damage estimation.



In this paper, we introduce a Science Gateway\changemarker{-based} framework designed for \changemarker{the creation of portable and reusable} post-event analysis pipelines developed in \cite{imbrosciano2025}. \changemarker{The principal scientific contribution consists in the migration and integration of validated but standalone workflows--based on Photogrammetry, AI and Data Visualization--into a single orchestrated pipeline within a Science Gateway platform,} to assess extreme natural events and analyze their effects on assets at risk.

The remaining parts of this paper are structured as follows. The next section describes the relevant related works. 
Section~\ref{gateway_platform} describes the science gateway platform, introducing the storage and computing infrastructure, including its architecture and workflow orchestration through Apache Airflow. Section~\ref{post-event} outlines the overall post-event analysis pipeline of our proposed approach, from UAV images acquisition to web application. Section~\ref{wf_implementation} describes the workflow implementation on Directed Acyclic Graphs (DAGs) to manage and automate the photogrammetry and machine learning-based processes. 
Section~\ref{results} shows our preliminary results. Section~\ref{conclusion} draws our conclusions and future works.

\section{Related Works}
\label{related work}

CyberGISX\footnote{CyberGISX: \url{https://cybergisxhub.cigi.illinois.edu/}} \cite{yin2019cybergis} provides an integrated platform for the development and sharing of geospatial software, with robust support for CyberGIS frameworks \cite{wang2016open}, geovisual analytics, spatial data processing, simulation, and applications across diverse domains such as agriculture, geography, health, and hydrology. Complementing this, the CyberGIS-Compute framework\footnote{CyberGIS-Compute: \url{https://github.com/cybergis/cybergis-compute-core}} offers a middleware solution that enables seamless integration with high-performance computing (HPC) resources through a Python-based software development kit (SDK). It supports large file transfers via Globus, facilitates community-driven model sharing via GitHub, and simplifies HPC job submission through a user-centric SDK interface. Furthermore, it provides fine-grained control for Slurm-based workflows, making it a versatile tool for advanced geospatial computation.

MyGEOHub\footnote{MyGEOHub: \url{https://mygeohub.org/}} \cite{kalyanam2019mygeohub} is a science gateway built on the HUBzero cyberinfrastructure framework \cite{mclennan2010hubzero}, offering integrated geospatial data services and computational tools. It serves a range of interdisciplinary research and education initiatives, particularly in hydrology, climate science, and agricultural economics, by providing a collaborative environment for sharing models, data, and analysis tools.

Quakeworx\footnote{Quakeworx: \url{https://quakeworx.org/}} \cite{chourasia_2024_13864099} is an emerging platform focused on the seismic science community, providing access to advanced rupture forecast models and validation tools. Built on the OneSciencePlace platform\footnote{OneSciencePlace: \url{https://onescienceplace.org/}} \cite{benham2019hubzero}, it embodies a composable and content-centric approach to cyberinfrastructure. OneSciencePlace supports FAIR (Findable, Accessible, Interoperable, Reusable) data and computing principles, promoting rapid deployment of scientific platforms, community engagement, and measurable impact at reduced cost.


These science gateways \changemarker{represent} mature \changemarker{science gateway and} cyberinfrastructure platforms. They \changemarker{typically offer integrated, domain-specific environments, providing user-friendly access to computational resources, collaborative tools, and pre-defined workflow mechanisms.
In contrast, our work takes a distinct approach by exploring the implementation of scientific processes across different modern workflow paradigms. We investigate the application of Apache Airflow for programmatic orchestration, Common Workflow Language (CWL) for declarative and portable workflow definitions, and Docker for ensuring reproducible execution environments. Such versatility allows these individual technologies to offer superior flexibility, fine-grained control, and guaranteed reproducibility for complex scientific computations.} 

\section{Science Gateway Platform}
\label{gateway_platform}
\subsection{Data and Computing Infrastructure}
The Science Gateway relies on the INAF PLEIADI\footnote{PLEIADI: \url{https://pleiadi.readthedocs.io/en/latest/}} infrastructure, leveraging its distributed computing capabilities to orchestrate and execute complex data processing workflows.

PLEIADI is a project led by USC VIII-Computing of INAF (National Institute for Astrophysics), that provides
high-performance computing (HPC) and high-throughput computing (HTC) resources.
In line with the standard model of Science Gateway infrastructures, individual researchers and teams involved in research projects - including
European projects, PRIN, INAF mainstream projects, and scientific missions -
can apply for access to its computing resources.  

The PLEIADI infrastructure is distributed 
across three main Italian sites:
 Bologna, Catania, and Trieste. In particular, Catania site hosts
56 computing nodes of which 18 with 256GB and 38 with 128GB of RAM. 
The configuration of each node is reported in Table \ref{tab:node-pleiadi}.

\begin{table}[ht]
    \centering
    \begin{tabular}{|c|c|}
        \hline
        Architecture & x86  \\
        \hline
         CPU Model & Intel(R) Xeon(R) CPU E5-2697 v4 @ 2.30GHz \\
         \hline
        Number of CPUs & 36 \\
         \hline
        Socket(s) & 2 \\
         \hline
        Core(s) per socket & 18 \\
         \hline
        Thread(s) per core & 1 \\
         \hline
    \end{tabular}
    \caption{Single computing node specifications - Pleiadi Catania}
    \label{tab:node-pleiadi}
\end{table}

The Catania site offers 174 TB of storage with BEEGFS\footnote{BEEGFS: https://www.beegfs.io/c/}, a parallel file system. Thanks to the \textit{Omni-Path HFI Silicon 100 Series, 100 Gbit interconnect}, the high-performance interconnectivity is managed in a highly efficient and scalable manner.
Each site includes its own front-end node that allows users to submit jobs, and the submission is managed by SLURM\footnote{SLURM: https://slurm.schedmd.com/documentation.html} job scheduler. 
This setup enables multiple users to run workloads concurrently,
with fair scheduling and resource limits that prevent a single user from consuming all the resources or interrupting the others' workloads.
Since March 2025, the Catania site has been expanded
with 10 new GPU nodes, featuring the  specification reported in Table \ref{tab:gpu-pleiadi}.

\begin{table}[ht]
    \centering
    \begin{tabular}{|c|c|}
         \hline
        Architecture & ppc64le \\
         \hline
        CPU Model & POWER9 2.1 \\
         \hline
        Number of CPUs & 128 \\
         \hline
        Socket(s) & 2 \\
        \hline
        Core(s) per socket & 16 \\
        \hline
        Thread(s) per core & 4 \\
        \hline
        GPU & 4 (TESLA V100) \\
        \hline
    \end{tabular}
    \caption{Single GPU node specifications - Pleiadi Catania}
    \label{tab:gpu-pleiadi}
\end{table}

\subsection{Science Gateway}

The Science Gateway is based on the Apache Airflow\footnote{Apache Airflow: \url{https://airflow.apache.org/}} workflow platform. 
This open-source 
infrastructure
is aimed to develop, schedule, and monitor batch-oriented
or event-driven
workflows \cite{harenslak2021data}. 


In Airflow, such processes are constructed using Python scripts, which outline Directed Acyclic Graphs (DAGs). These graphs are “directed” - indicating a predetermined order of task execution, and “acyclic” - guaranteeing the absence of cycles or loops within the workflow. In each DAG script, the tasks are represented as nodes, with the dependencies between these tasks depicted as edges. The most commonly used approach for implementing tasks is through operators, as they offer a versatile and straightforward method for defining various types of tasks. Operators encapsulate the logic necessary to perform specific actions, such as running a Python function, executing a Bash command, transferring data between systems, or interacting with external services like databases or cloud platforms. Besides operators, Airflow also offers alternative approaches such as the TaskFlow API\footnote{TaskFlow API: \url{https://airflow.apache.org/docs/apache-airflow/stable/core-concepts/taskflow.html}} or custom operator implementation for more specialized requirements.

Airflow’s architecture is composed of a scheduler, a web server, a metadata database, an executor, and workers. We briefly discuss each element below:

\begin{itemize}
   \item Metadata Database: Airflow uses a metadata database to store information about DAGs, tasks, and their status. This allows for tracking and managing the state of workflows.
   \item Web Server: Airflow provides a web-based user interface to monitor and manage workflows. Users can visualize DAGs, view task logs, and manually trigger or pause DAG runs.
   \item Scheduler: The scheduler monitors all tasks and DAGs to decide what needs to be run, based on dependencies and schedules. It sends tasks to be executed to the executor. Different types of executors are available in Airflow including the \textit{LocalExecutor}, which handles tasks locally for parallel processing, and the \textit{SequentialExecutor}, ideal for sequential task execution in simpler setups. For distributed task execution, Airflow provides the \textit{CeleryExecutor}, utilizing Celery for task distribution across multiple machines, and the \textit{KubernetesExecutor}, which scales tasks by creating separate pods within a Kubernetes cluster. In environments designed for distributed and parallel execution, tasks are allocated across various workers, i.e., the computational units responsible for executing tasks, enabling them to be executed simultaneously.
   \item A folder of DAG files: DAGs are typically stored as Python files in a designated folder. The scheduler monitors this folder, reading the DAG files to understand the task dependencies and execution schedules, 
   ensuring the smooth execution of workflows.
\end{itemize}

\begin{figure}
    \centering
    \includegraphics[width=\linewidth]{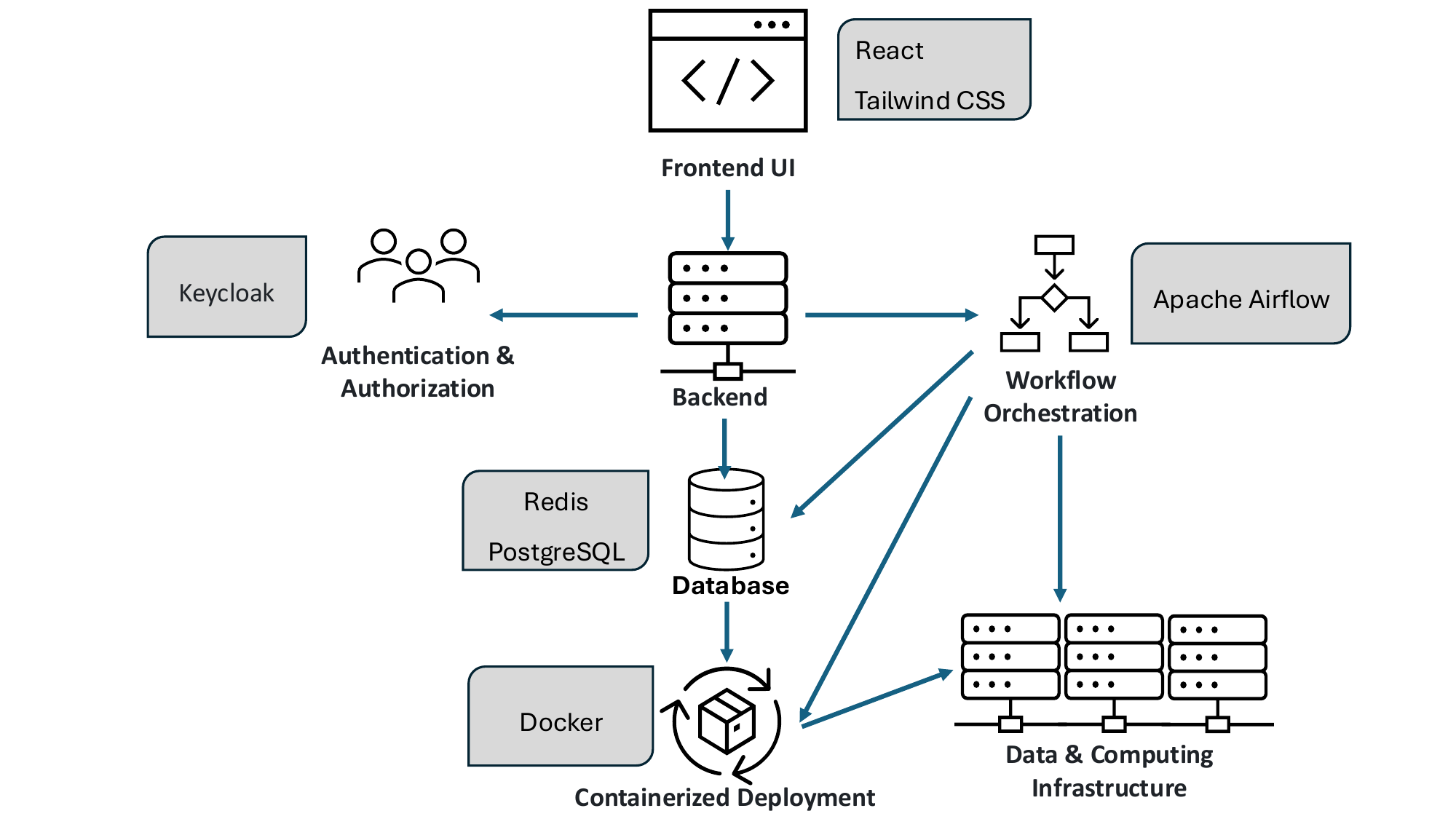}
    \caption{Science Gateway distributed deployment based on the Workflow platform}
    \label{fig:sg}
\end{figure}

We opted for a distributed deployment of the Science Gateway to offer scalability and reliability by separating its key components across different servers aiming to improve
performances and system maintainability.

The main components include, as shown in Fig. \ref{fig:sg}: a Frontend UI based on React for users to define and monitor workflows; Workflow Orchestration with Apache Airflow for scheduling and executing jobs; Authentication \& Authorization with Keycloak for user management and authentication; a FastAPI Backend for communication between the frontend, Keycloak, and the Workflow orchestration; and, finally all components run in Docker containers for easy setup and management.

\section{Post-event analysis pipeline}
\label{post-event}
Natural disasters have unpredictable effects on assets and infrastructure, making post-event analysis a challenging but essential task for insurance companies and public administrations.
Due to the absence of standardized approaches capable of addressing the diverse impacts of environmental disasters, the science gateway aims to provide methodologies for damage evaluation that yield results applicable regardless of the specific natural disaster being analyzed.
A primary objective of this activity is to streamline damage assessment processes, reducing the need for time-intensive on-site inspections. 

\begin{figure}[!ht]
    \centering
    \includegraphics[width=0.8\linewidth]{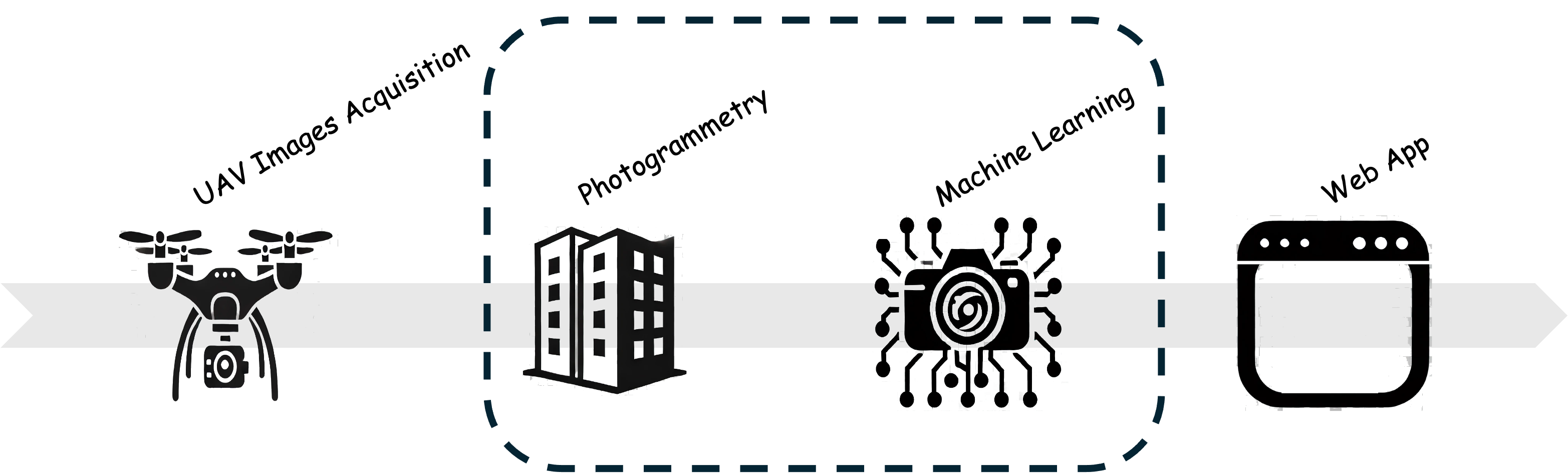}
    \caption{Post-event analysis pipeline}
    \label{fig:pipeline}
\end{figure}

The pipeline involves the following steps depicted in Figure \ref{fig:pipeline} (see also \cite{imbrosciano2025} for more details):

\begin{enumerate}
    \item \textbf{UAV Images Acquisition} 
    To ensure both effectiveness and precision in the final results, the initial phase involves planning and execution of drone surveys over the affected area. This process includes delineating the area of interest and setting drone flight specifics—such as altitude, flight path, and coverage area.
    
    \item \textbf{Photogrammetry} The collected imagery is then used to generate high-resolution (centimeter-scale) 3D models using the ``Aerial Structure-from-Motion" (ASfM) techniques~\cite{aguera2017accuracy,knuth2023historical}. 
    This photogrammetric approach autonomously determines the geometry of the area, as well as the position and orientation of the cameras.  
    By leveraging overlapping images captured from multiple viewpoints, it enables the reconstruction of detailed 3D
    models and produce georeferenced maps of the examined area.
     
    \item \textbf{Machine Learning} The process is complemented by Artificial Intelligence (AI) algorithms that extract features from aerial images and integrate them into the model, enriching it with semantic information.
    \item \textbf{Web App} These steps aim to detect and classify damages in 3D models, offering a scalable framework for post-disaster analysis. The dataset of augmented digital twins models provides stakeholders and claims adjusters with detailed visual references for remote damage assessment and is accessible through a user-friendly web application. 
    \changemarker{The platform allows for the immediate 3D visualization of the analysis results, powered by Cesium\footnote{CesiumJS: \url{https://cesium.com/platform/cesiumjs/}}, facilitating a prompt understanding of the damage location and extent. Furthermore, the resulting data are downloadable, allowing further offline analysis and integration into other systems.}
\end{enumerate}

In this work we specifically focused on step 2) and 3) detailed in the next Section \ref{wf_implementation}.

\section{Workflow Implementation}
\label{wf_implementation}

\begin{figure*}[htp]
    \centering
    \includegraphics[width=0.95\linewidth]{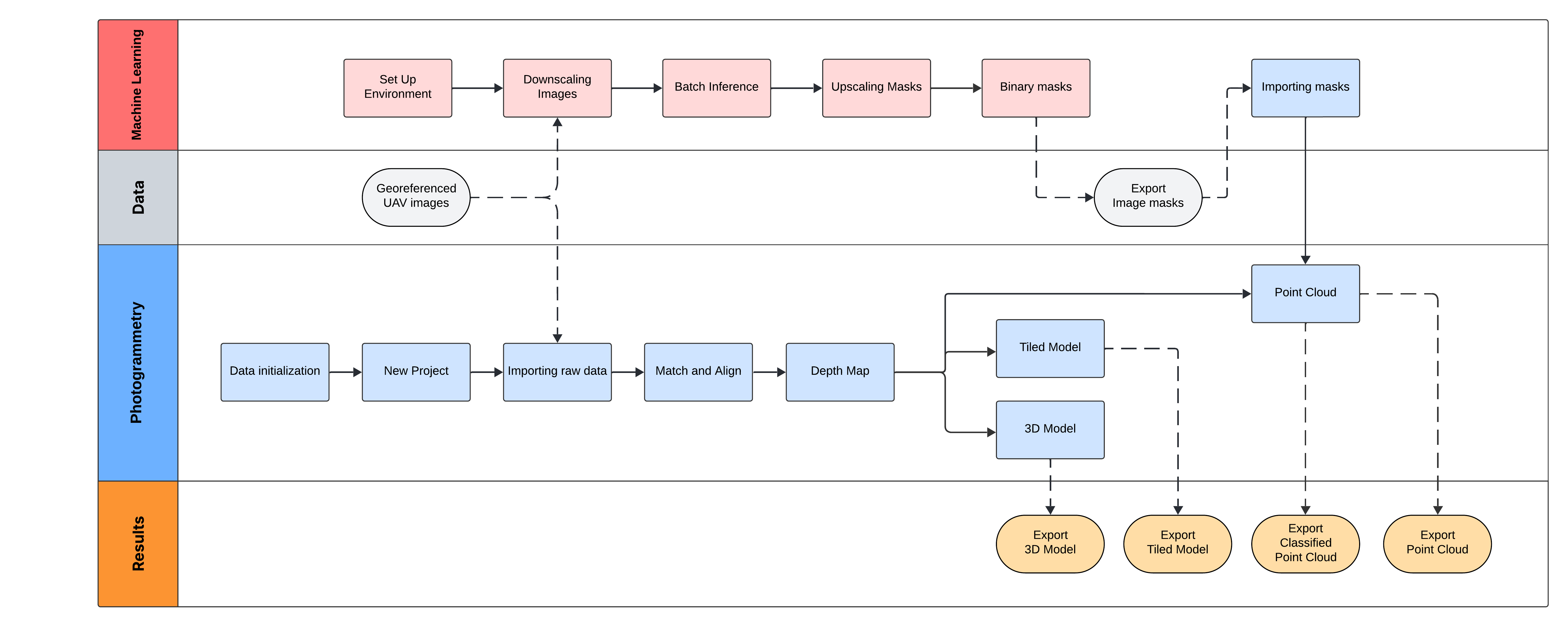} 
    \caption{Graphical overview of the workflow and exported data}
    \label{fig:airflow-dag}
\end{figure*}

We have implemented\footnote{https://github.com/VisIVOLab/Post-Event-Analysis-Workflow}, as shown in Figure~\ref{fig:airflow-dag}, two distinct DAGs, one for the photogrammetric process (in light blue) and another one for the machine learning process (in light red). 
\changemarker{While these were consistently implemented using CWL, Docker, and Airflow, we will discuss the Airflow version in detail here;}
the remaining notations represents the input data (in light gray) and the resulting output (in light yellow). Each DAG is organized using the Docker, Bash and Python operators provided by Airflow, ensuring a clear and modular task structure by limiting dependencies and configurations to the individual task level.

Moreover, the overall workflow adopts an ETL-based structure: UAV imagery is extracted and transformed through photogrammetric processing, \changemarker{then semantically segmented using a machine learning model,
ultimately producing 3D data models.}

The entire photogrammetric and machine learning processes—from the import of raw images to the generation and refinement of 3D models—has been segmented into individual tasks to align with the structure of DAG's Airflow.

\subsection{Photogrammetric DAG process}

The initial step executes the configuration of the working environment. It is responsible for receiving the setup defined by the end user and adapting the development environment with these settings. The front-end interface generates a custom configuration file named~\textit{dagrun.cfg}, which encapsulates all the parameters and execution preferences required for the photogrammetric workflow process.
The file serves as a consistent and centralized input reference for all photogrammetric tasks, ensuring traceability and reproducibility throughout the structure from motion workflow.


These settings allow the initialization of a new project and the start of the photogrammetric process using the proprietary software Agisoft Metashape\footnote{Agisoft Metashape. Available online: \url{https://www.agisoft.com/}}, version 2.2.0 -- the latest available release at the time of writing. Its Python module is highly versatile and specifically designed to create Digital Twins that are both georeferenced and to scale, making them suitable for precise measurements and analyses within Geographic Information System (GIS) environments. These features make the outputs particularly valuable for planning, design, and spatial studies, especially in hazard mapping contests.

Once the images have been imported, preprocessing begins by filtering them, with particular attention to exclude blurry, overexposed, underexposed images; which ensures that only high-quality images suit the 3D reconstruction.

The images were intentionally acquired with a high degree of overlap to facilitate the subsequent image matching and alignment processes during photogrammetric reconstruction. These steps are based on the extraction of distinctive features from the images, which serve as key reference points to estimate the camera position for each photo.

To understand the 3D structure of the environment, the next task generates the depth maps. This process consists in estimating the distance of each pixel in all images from the cameras, which gives the possibility to represent the scene in terms of depth information.

Depending on the photogrammetric DAG selected, it is possible to proceed in two different ways. One approach involves directly generating the point cloud, tiled model, and 3D model from the depth maps — as shown in Figure~\ref{fig:airflow-dag} this is generally considered the most effective method in terms of output quality, it supports GPU acceleration and mostly provides better results for objects and scenes with a large number of minor details. Alternatively, it is possible to first compute the point cloud and then use it as the source for generating the tiled model and the 3D model based on the wireframe, followed by texture mapping to enhance realism. In the end, based on the configuration defined in the \textit{dagrun.cfg} file, it is possible to export the obtained results in the preferred formats and resolutions.


\subsection{Machine Learning DAG process}
\label{subsection:ml}

The second DAG focuses on semantic segmentation using deep learning. Starting from UAV images destined to the photogrammetric process, the DAG generates black-and-white PNG masks that highlight the shape of selected object classes. These masks are then used by Metashape to classify the corresponding objects within the 3D model.

An initial setup task clones the machine learning repository~\footnote{https://github.com/ICSC-Spoke3/HaMMon-ML-digital-twin/} where the scripts are stored. 
This step is useful either when running directly 
on a local machine or in containerized environments, where the repository folder can be mounted across task-executing containers after a single pull operation.
Once the environment is ready, the pipeline begins by downscaling input images to a resolution consistent with the model’s training datasets: this helps to limit memory requirements and is expected to improve inference quality. 

All processing steps on images and labels are carried out by Python scripts, 
which can be executed in containers or triggered using BashOperators. 
These scripts leverage the multiprocessing module to parallelize the work 
across all available CPU cores, processing different images concurrently.
Intermediate artifacts—such as temporary PNG files and JSONs 
containing metadata—are stored in temporary directories 
to be accessed by downstream tasks and automatically removed at the end
of the workflow to maintain a clean execution environment.

Following the initial resizing step, a dedicated Python module is launched to run inference 
in batches on the downscaled images using GPU acceleration, leveraging PyTorch 
vectorization for efficient computation. The deep learning model currently used is ``100 Layers Tiramisu"~\cite{Jegou2017} a Fully Convolutional neural network, trained on datasets such as FloodNet~\cite{FloodNet2021} and RescueNet~\cite{RescueNet2023}. These datasets provide UAV imagery captured after hurricanes, highlighting both environmental features (e.g., roads, trees, buildings) and disaster-specific elements (e.g., debris, flooded areas).

Inference produces grayscale segmentation masks in which 
each pixel encodes a class label. These masks undergo formal validation and are then upscaled to match the original image resolution. While a simple nearest-neighbor method is currently used, more sophisticated deep learning-based upscaling methods (which take into account features from the original images) are planned for production environments.

Finally, each mask is converted into a set of binary black-and-white masks—one per class—suitable for ingestion into photogrammetric software like Metashape. 
This approach enables seamless integration between 2D semantic data and 3D reconstruction workflows.

\section{Results}
\label{results}

This section presents the preliminary results achieved in the post-event analysis.
As part of the HaMMon project, INAF-OACT conducted a field survey in October 2024 within the municipality of Tredozio
province of Forlì-Cesena (Emilia-Romagna, Italy), to document and analyze the impacts of landslides, seismic activity, and flooding caused by the Tramazzo river. 
The operation was supported by equipment and drones, which play a crucial role in hazard mapping over the past years~\cite{kucharczyk2021remote, daud2022applications}, useful to address challenges in disaster management and risk assessment, especially in areas that are difficult or dangerous to access.



\begin{figure*}[t]
    \centering
    \begin{minipage}{0.47\linewidth}
        \centering
        \includegraphics[width=\linewidth]{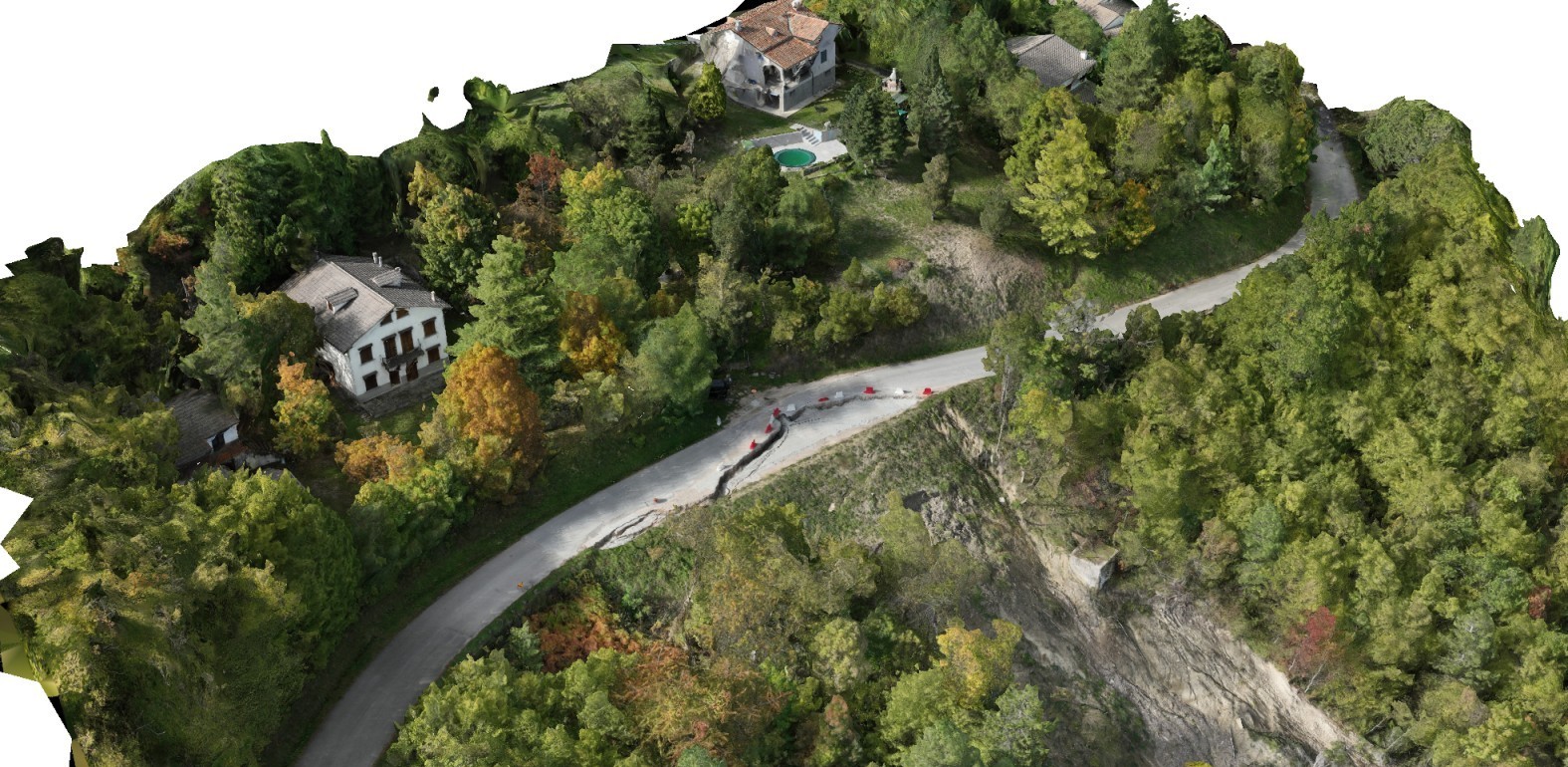}
        \caption*{Tiled Model}
    \end{minipage}
    \hfill
    \begin{minipage}{0.47\linewidth}
        \centering
        \includegraphics[width=\linewidth]{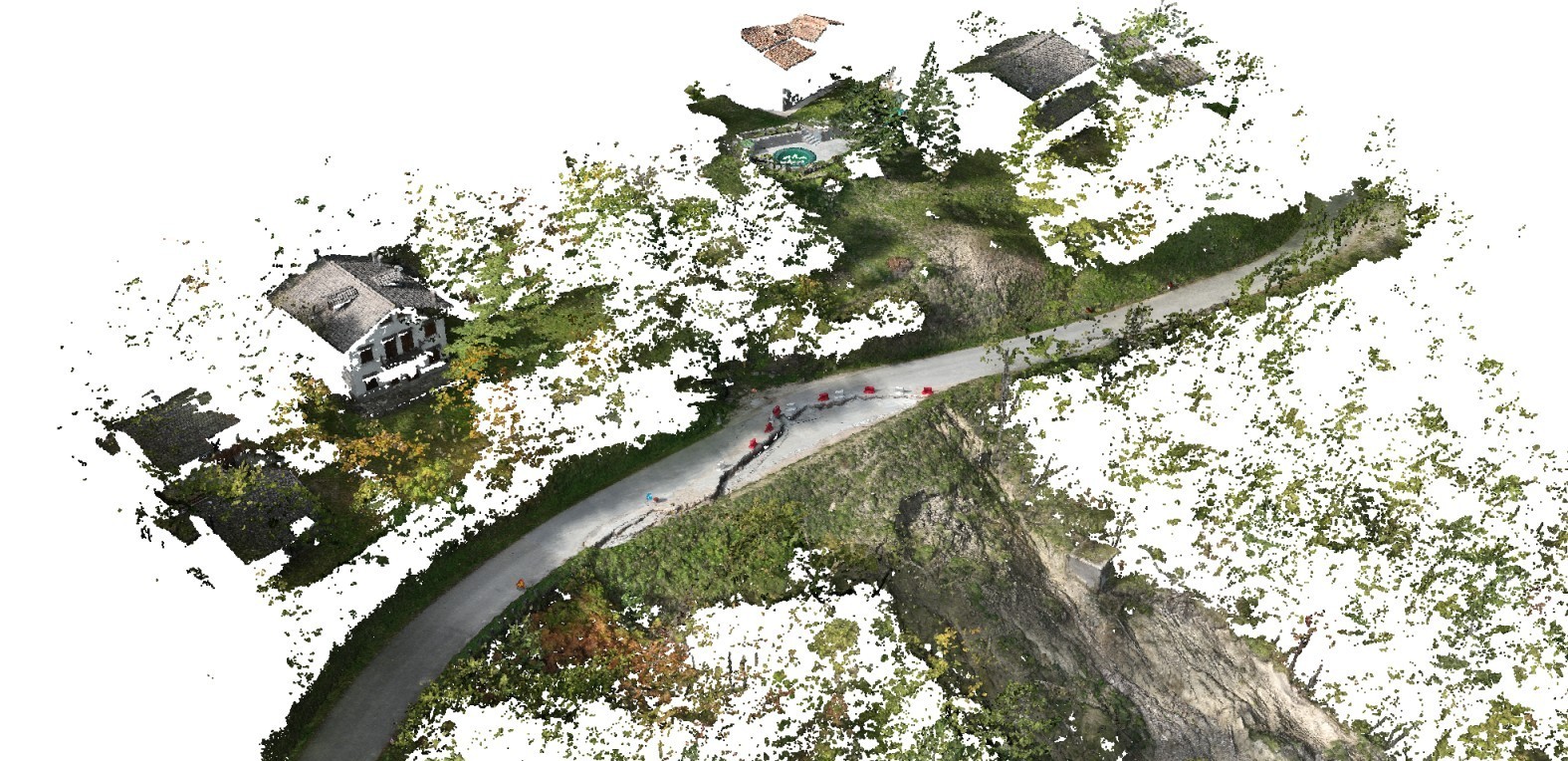}
        \caption*{Colored Point Cloud}
    \end{minipage}
    
    \vspace{2mm}
    
    \begin{minipage}{0.48\linewidth}
        \centering
        \includegraphics[width=\linewidth]{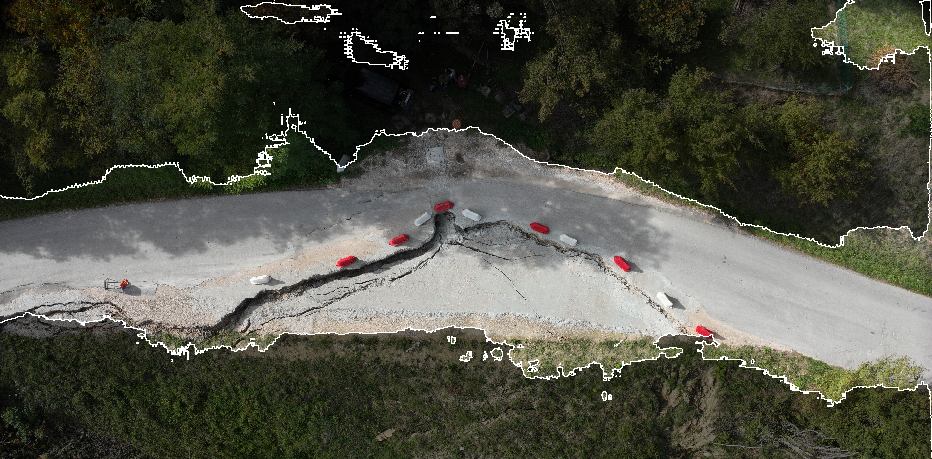}
        \caption*{Example of a segmentation mask generated through machine learning, highlighting the road area}
    \end{minipage}
    \begin{minipage}{0.48\linewidth}
        \centering
        \includegraphics[width=\linewidth]{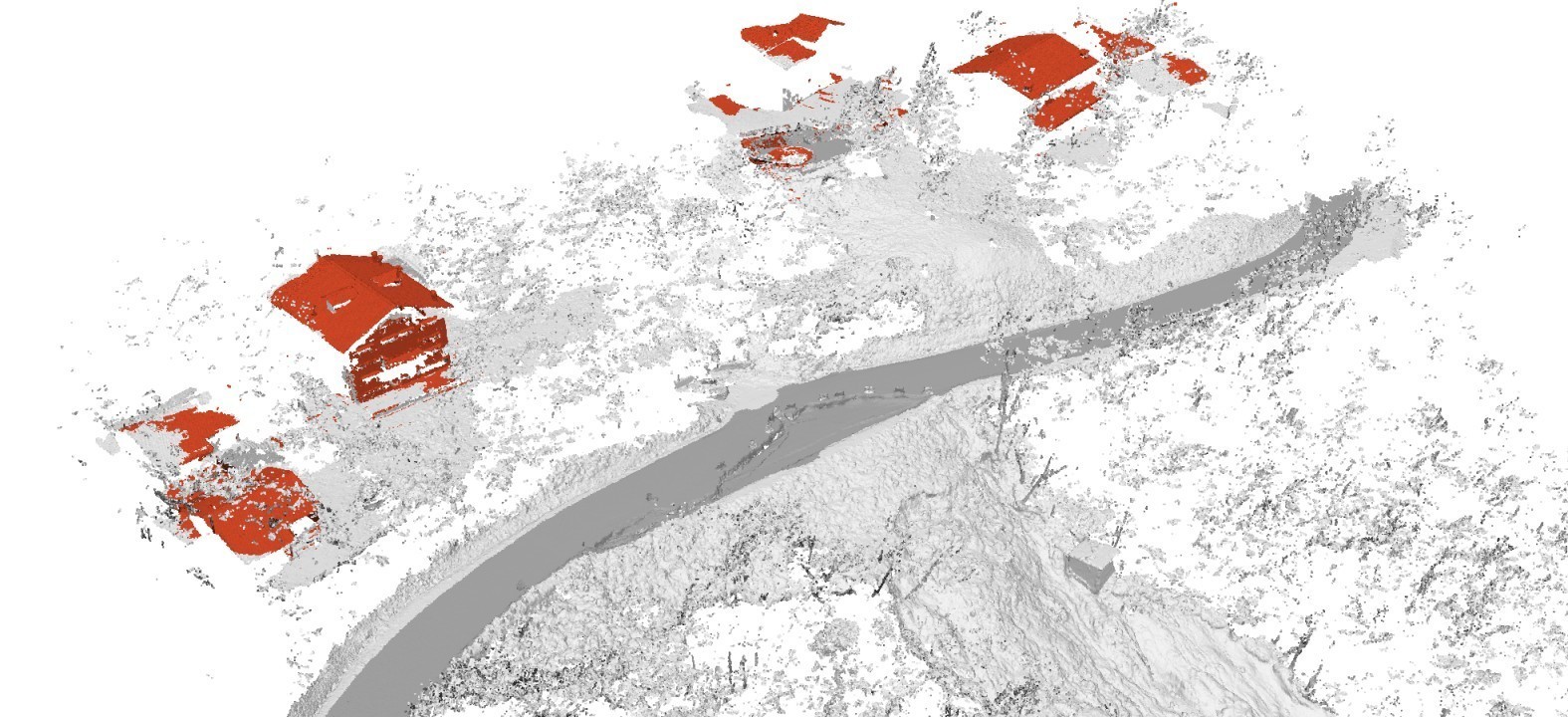}
        \caption*{Point cloud classification of buildings (red) and roads (grey)}
    \end{minipage}
    \hfill
    
    \caption{Visual outputs generated by the science gateway on the survey carried out in Tredozio (Monte Busca area)}
    \label{fig:output}
\end{figure*}

Data acquired on-site in Monte Busca area were processed using the science gateway platform. The main results generated through this workflow are summarized in the Figure~\ref{fig:output} which illustrates the results obtained. The top-left quadrant shows the high-detail tiled 3D model. The model clearly reveals the impact of the landslide. The affected area was estimated to be approximately 30 meters in width and 110 meters in depth. The top-right quadrant displays the corresponding colorized dense point cloud. The result lacks rich detail due to the presence of dense vegetation, which introduces irregularities and noise in the photogrammetric reconstruction process.
The images below present the preliminary results obtained through the integration of the Artificial Intelligence classification module. On the left, a segmentation result is overlaid on the original UAV image; on the right, the classified point cloud is shown, obtained by projecting all computed masks from the input images onto the 3D model.
The results are noticeably affected by the fact that the input images differ significantly from the training domain, both in terms of represented objects and image characteristics such as angle and distance. The development is still at an early stage, and more robust models with improved generalization capabilities will be employed in future iterations. Nevertheless, these outcomes offer a solid starting point for future development and serve as a valuable test for assessing the integration and coherence of the automated workflow.


\section{Conclusions}
\label{conclusion}
In this work, we presented a structured 
\changemarker{workflow} for the post-event analysis, integrating photogrammetric processing and machine learning-based semantic segmentation, all orchestrated through a science gateway platform on Pleiadi infrastructure.

The preliminary results obtained from the field survey conducted in the Monte Busca area (Tredozio) demonstrate the potential of the proposed approach to assess extreme natural events and analyze their effects on assets at risk.


Future work will focus on the refinement of the processing workflows to enhance accuracy and scalability within the science gateway infrastructure. 
\changemarker{In addition, we acknowledge the importance of complementing our qualitative findings with a through measurable evaluation. As this work represents preliminary validation results, we plan to conduct a thorough quantitative analysis in future developments, to assess overall performance and scalability of the implemented workflows.}

\section*{Acknowledgment}
The work is supported by the Spoke 1 ``FutureHPC \& BigData'' and the Spoke 3 ``Astrophysics and Cosmos Observations'' of the ICSC – Centro Nazionale di Ricerca in High Performance Computing, Big Data and Quantum Computing, funded by NextGenerationEU.



%
\bibliographystyle{IEEEtran}
\bibliography{main}

\end{document}